\documentclass[a4paper]{jpconf}
\usepackage{graphicx}
\begin{document}
\title{Air Shower Measurements in the Primary Energy Range from 
PeV to EeV}

\author{Andreas Haungs}

\address{$^\ast$Institut\ f\"ur Kernphysik, Forschungszentrum Karlsruhe,
76021~Karlsruhe, Germany}

\ead{haungs@ik.fzk.de}

\begin{abstract}
Recent results of advanced experiments with sophisticated measurements 
of cosmic rays in the energy range of the so called knee 
at a few PeV indicate a distinct knee in the energy spectra 
of light primary cosmic rays and an increasing dominance of heavy 
ones towards higher energies. 
This leads to the expectation of knee-like features 
of the heavy primaries at around 100 PeV. To investigate 
in detail this energy 
region several new experiments are or will be devised.

\end{abstract}

\section{Introduction}

\begin{figure*}[ht]
\centering
%\vspace*{0.1cm}
\includegraphics[width=155mm]{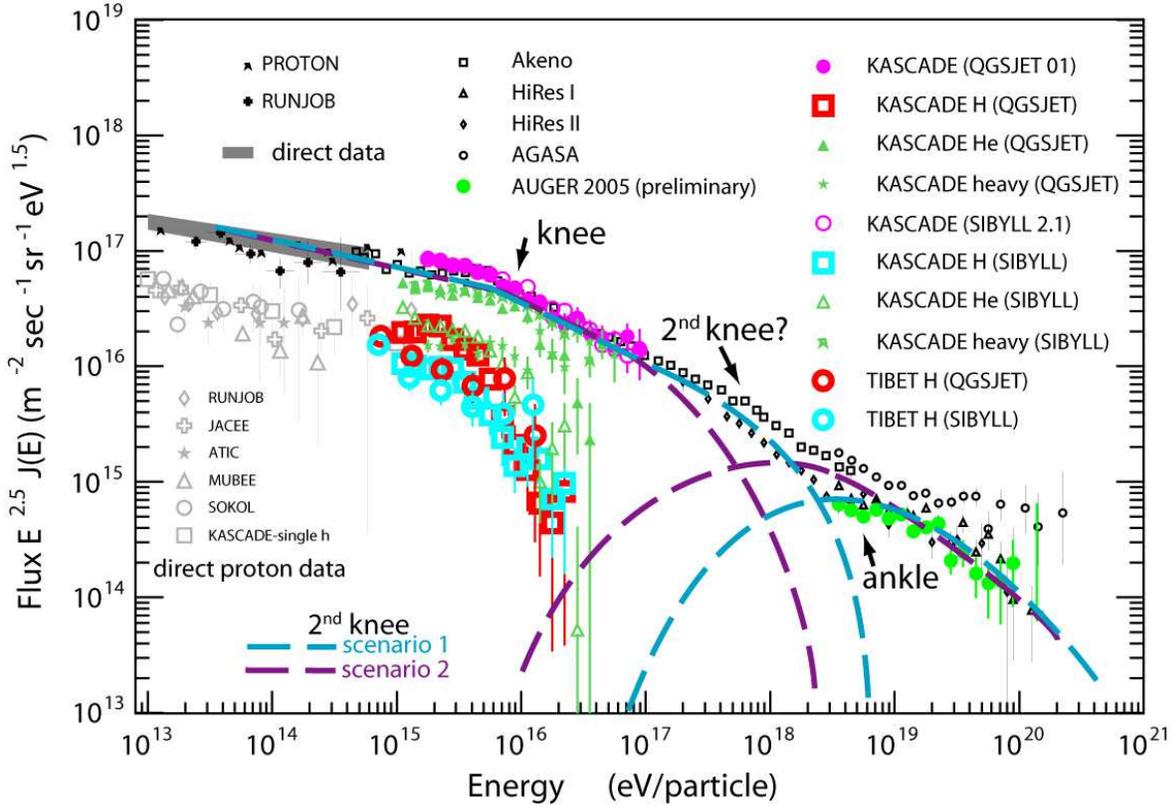}
\caption{\label{spectrum}Primary cosmic ray flux. 
Results of some experiments for the all-particle spectrum as well as 
for spectra of individual mass groups, in particular for protons. 
Possible scenarios for the origin of a second knee 
are also shown (see text).}
\end{figure*}

In the primary energy range of PeV to EeV direct
measurements of cosmic rays are presently hardly possible 
due to the low flux, but 
indirect measurements observing extensive air showers (EAS) are 
performed. The all-particle energy 
spectrum in this range shows one distinctive
discontinuity at few PeV, known as the knee, 
where the spectral index of a power-law dependence changes from $-2.7$ to approximately 
$-3.1$. In addition, at a few hundred PeV, there 
are weak experimental hints for a further, much less pronounced, change of the 
the power law index, assigned as so called second 
knee~(see Fig.~\ref{spectrum}).  

Astrophysical scenarios like the change of the acceleration 
mechanisms at the cosmic ray sources 
(supernova remnants, pulsars, etc.) 
or effects of the transport mechanisms inside the Galaxy (diffusion 
with escape probabilities) are conceivable for the origin of the knee
as well as particle physics reasons like a new kind of hadronic 
interaction inside the atmosphere or during the transport through 
the interstellar medium. 
The highest energies above the so called ankle at a few EeV
are believed to be exclusively of extragalactic origin. 
Thus, in the experimentally scarcely explored region between 
the first (proton) knee and the ankle there are two more 
peculiarities of the cosmic ray spectrum expected: 
(i) A knee of the heavy component which is 
either expected (depending on the model)
at the position of the first knee scaled with Z (the charge) 
or alternatively with A (the mass) of iron. 
(ii) A transition region from galactic to extragalactic origin 
of cosmic rays, where there is no theoretical reason for a 
smooth crossover in slope and flux. Dependent on the considered 
astrophysical model the second knee is allocated to case (i) or
(ii), respectively.   

It is obvious that only detailed measurements over the whole energy
range from $10^{14}\,$eV to $10^{18}\,$eV
and the reconstruction of individual primary energy spectra for the 
different incoming 
particle types can validate or disprove some of these models. 

Despite EAS measurements with various different 
experimental setups in the 
last decades this demand could never accomplished, mainly due to 
the weak mass resolution of the measured shower 
observables~\cite{rpp}. 

Recently, however, there are a few experiments which could 
reconstruct individual energy spectra for different mass groups 
in the knee region (KASCADE, Tibet-AS$\gamma$, EAS-TOP) by accurate
hybrid air-shower measurements. 
And there are promising experiments in set-up or
proposed (KASCADE-Grande, IceTop, Pierre-Auger-Observatory-, and
Telescope Array - enhancements) which will
provide a similar accuracy for measurements in 
the energy range from 50-1000 PeV.

\section{The (first) knee}

The most interesting results in analyzing EAS in the knee 
region are originating  
from the KASCADE experiment, where the data analyses aim to 
reconstruct the energy spectra 
of individual mass groups, taking into account not only different
shower observables (muon and electron shower sizes), 
but also their correlations on an event-by-event basis.
By applying unfolding procedures (based on Monte-Carlo simulations
using different hadronic interaction models) to the experimental 
data individual energy spectra are obtained as displayed in 
Fig.~\ref{spectrum}. 
A knee like feature is clearly visible in the all particle spectrum,
which is the sum of the unfolded single mass group spectra, 
as well as in the spectra of primary proton and helium.
KASCADE claims that the elemental composition of cosmic
rays is dominated by the light components below the knee and 
by a heavy component above the knee feature~\cite{Ulric05}.
\begin{figure}[ht]
\begin{minipage}{18pc}
\centering
\includegraphics[width=18pc]{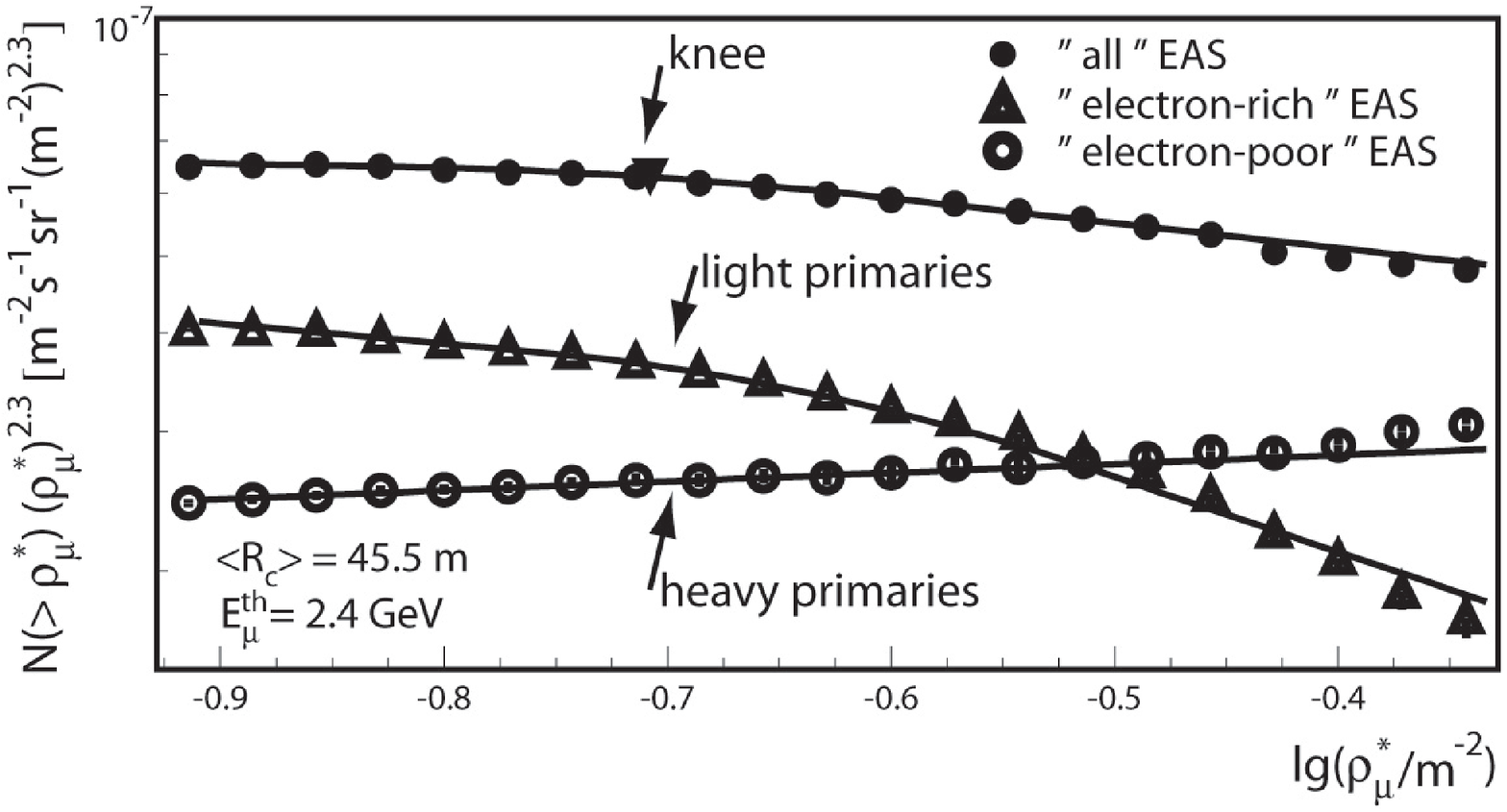}
\caption{\label{muon}Muon density spectra for different samples of EAS
as obtained by KASCADE measurements.}
\end{minipage}\hspace{20pt}%
\begin{minipage}{18pc}
\centering
\includegraphics[width=18pc]{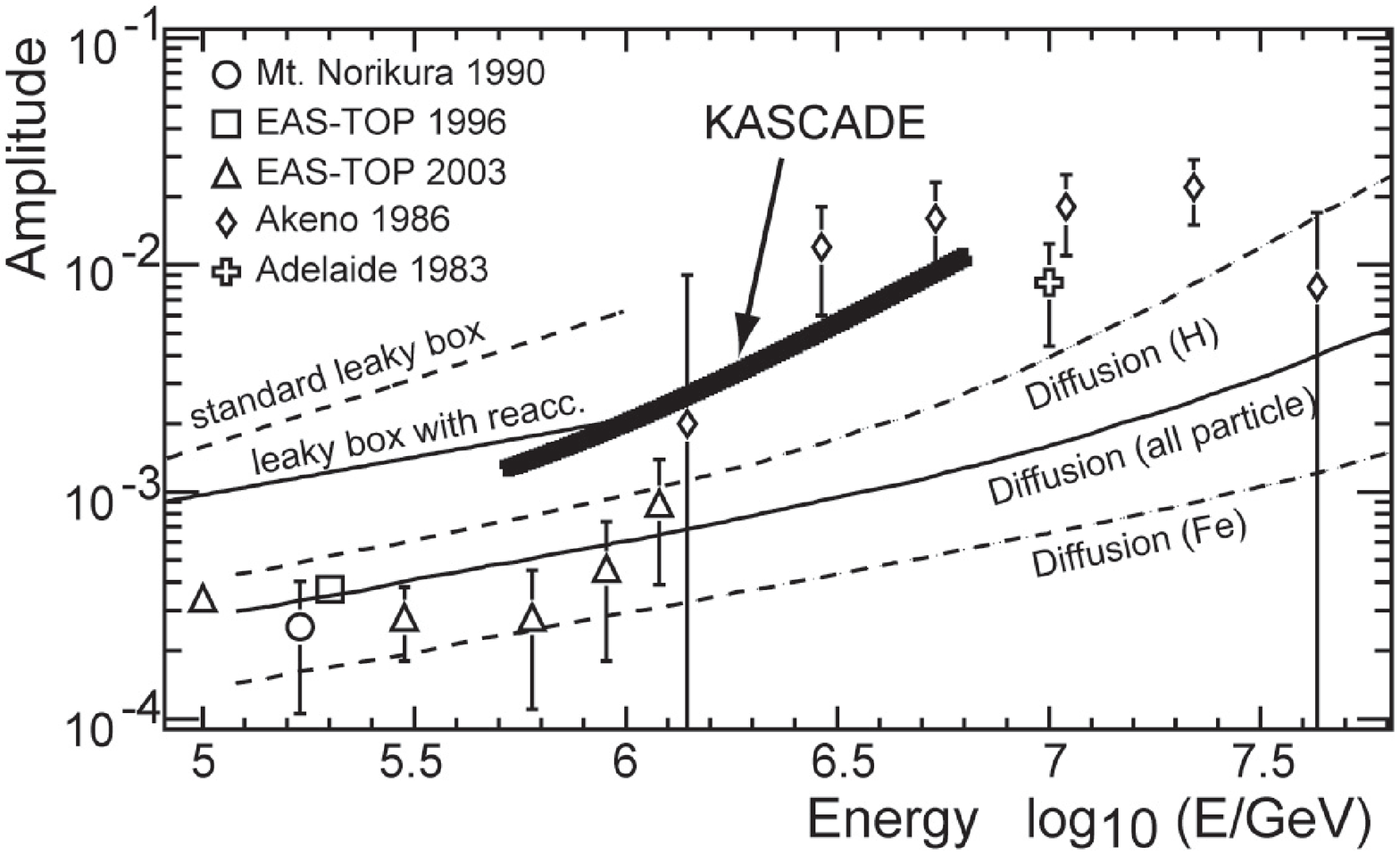}
\caption{\label{aniso}Rayleigh amplitude of the harmonic analyses 
of the KASCADE data (limit on a 95\% 
confidence level).}
\end{minipage} 
\end{figure}

Data of the EAS-TOP air-shower experiment were also analyzed to 
obtain energy spectra of individual mass groups 
(of light, medium, and heavy primaries in case of
correlating shower size with low energy muon densities~\cite{EAS}, 
and of light and heavy in case of correlating shower size with high 
energy muons above 1 TeV measured by the MACRO detector~\cite{MACRO}).
The results confirm the more detailed investigations by KASCADE in
the sense, that the knee is caused by the decreasing flux of the
light component.

In contradiction to the results of KASCADE and EAS-TOP are the
outcomes of the TIBET-AS$\gamma$ group~\cite{Tibet}.
Whereas the all-particle spectrum is obtained by using air shower
data alone which agree nicely with all other measurements, 
for getting individual mass spectra a different approach is chosen.
From coincident measurements of the air shower array with 
high-energy particle families in emulsion chambers
the spectra for primary protons and helium are extracted 
by using a neural net analysis. 
The small number of measured events (177 in 3 years), and the low 
detection efficiency, in particular for heavy elements, requires an
efficiency correction of up to factor 100 (which estimate is also
model dependent) to obtain the spectra as
plotted in Fig.~\ref{spectrum}.
Due to the fact that in the obtained proton spectrum no distinct
structure appears to be visible, but the index of an assumed power law 
differs from results of direct measurements, the TIBET group claims
that there must occur a proton knee at energies below
$10^{15}\,$eV. Hence the knee at a few PeV in the all-particle cosmic
ray spectrum must be dominated by a heavy component.
 
The largest uncertainty of all the discussed investigations is due to
differences in the hadronic interaction models underlying the
analyses. Modeling the hadronic 
interactions needs assumptions from particle physics theory and 
extrapolations resulting in large uncertainties, which are reflected 
by the discrepancies of the results presented here. 
These findings are confirmed by detailed investigations of further
shower observables measured by KASCADE~\cite{isv04}, in particular.
These investigations of observable correlations 
have shown that none of the present hadronic
interaction models is able to describe all the KASCADE data
consistently (on a level of a few percent). It was also found that
changing the hadronic interaction models do not change the form of
the individual energy spectra, but the relative abundances, only.
Nevertheless one should note that the uncertainties in the 
hadronic interaction model underlying the analyses prove to 
be the limit of the accuracy of the results.

Neglecting somehow the uncertainties by the interaction models, 
KASCADE used an additional independent and more direct approach to
interprete the measurements (Fig.~\ref{muon}). 
By using three observables (one observable as energy identifier - 
the local muon density of high energy muons; 
and two observables as mass identifier - the ratio of electron to 
low energy muon number for dividing the whole EAS sample in a sample 
generated by light primaries and heavy primaries) 
KASCADE could impressively and in a nearly 
model independent way demonstrate that the knee is caused by the 
decreasing flux of light primaries~\cite{muon}.   

In addition to the energy spectra of individual primaries, 
investigations of anisotropies in the arrival 
directions of the cosmic rays can give additional information on 
the cosmic ray origin and of their propagation.  
Depending on the model of the origin of the knee and on 
the assumed structure of the galactic magnetic field
one expects different amplitudes for the large-scale anisotropy 
in the energy range of the knee.
The limits of large-scale anisotropy analyzing the 
KASCADE data~\cite{Gmaier1} and other experiments
are shown in Fig.~\ref{aniso}. 
These limits already exclude particular model predictions. 

In summary, despite the recent success of sophisticated
experiments like KASCADE, EAS-TOP or TIBET there are still 
only weak constraints for distinguishing detailed 
astrophysical models explaining the knee in the
primary cosmic ray energy spectrum. In particular, due to the weak
mass resolution of the experiments, the question of mass or charge
dependence of the knee positions for the different primaries still remains 
open.

\section{The second knee}

To distinguish between astrophysical models of the origin of the
knee constraints can be given by clarifying the existence and 
source of the second knee, which is possible by determining  the mass
composition in the relevant energy range in detail. 
It is obvious that for the range 
between 50 and 1000 PeV  sophisticated experiments are needed 
to measure the EAS with the same statistical and reconstruction 
accuracy as the experiments discussed in the previous section to
distinguish between the two astrophysical scenarios shown in 
Figure~1.  
Presently, there are several efforts for that:

In 2003, the original KASCADE experiment was extended in 
area by a factor 10 to the new
experiment KASCADE-Grande~\cite{Navar04}.
KASCADE-Grande allows now a full coverage of the energy range 
around the knee, including the possible second knee.
KASCADE-Grande will be able to prove the existence of the knee 
corresponding to heavy elements.

At the south pole, the neutrino experiment IceCube has started its
deployment. Parallel to this under-ice experiment an air shower array
(IceTop) will be build on the surface. 
This setup will allow to measure in
coincidence the charge particles of air-showers and 
high-energy muons above 500 GeV with 
high accuracy. Therefore it will be able to reconstruct energy 
spectrum and mass composition in the energy range of the 
second knee~\cite{icetop}.

Finally, the Pierre-Auger-Observatory and Telescope-Array 
collaborations are discussing enhancements of their
hybrid detection techniques to cover also lower energies. 

\section{Conclusions}

From detailed EAS measurements at 1-10 PeV we do know that at a few 
times \mbox{10$^{15}$ eV} the knee is due to light 
elements, 
that the knee positions depend on the kind of the incoming particle, 
and that cosmic rays around the knee arrive our Earth 
isotropically.
New experiments, measuring higher energies, will be able to prove, 
if existent, the knee corresponding to heavy elements, 
They will give by far more constraints to the various astrophysical
models for explaining the origin and propagation of very high energy
cosmic rays.

\smallskip

\end{document}